\documentclass[prl,aps,showpacs]{revtex4}
\usepackage{graphicx}
\usepackage{dcolumn}
\usepackage{bm}
\usepackage{amsmath}
\begin{document}
\title{Kinetic phase diagrams of a ternary hard sphere mixture}
\author{A. Vizcarra-Rend\'on$^1$, A. Puga-Candelas$^1$, S. Aranda-Espinoza$^1$, M. A. Ch\'avez-Rojo$^2$, 
and R. Ju\'arez- Maldonado$^1$}
 \address{$^1$Unidad Acad\'emica de F\'{\i}sica, Universidad Aut\'{o}noma de Zacatecas,
Calzada Solidaridad Esquina con Paseo la Bufa S/N, C.P. 98060
Zacatecas, Zac., M\'{e}xico.\\
$^2$Facultad de Ciencias Qu\'imicas, Universidad Aut\'onoma de Chihuahua. Circuito No. 1, Nuevo Campus Universitario;
Chihuahua, Chuh. M\'exico.}
\date{\today}

\begin{abstract}
We use the Self Consisten Generalized Langevin Equation theory (SCGLE) 
to study the dynamic arrest transitions of a system of three species 
of hard sphere colloidal system in the size ratio
1:3:9. We find that
the inclusion of the smallest species has a depletion effect that
drives the system to a second glass-liquid-glass re-entrance 
in a similar way that the inclusion of a small species in an otherwise monodisperse
system leads to a (first) re-entrance. 
And we also find that the new glass after the second reentrance has very small localization length.
Additionally we compare the kinetic phase
diagram of a binary hard sphere mixture with size asymmetry 1:5, obtained with
Mode Coupling Theory (MCT) and SCGLE theory and exhibit the significant differences 
between the two theories..


\pacs 64.70.pv,64.70.P-
\end{abstract}
\maketitle
\section{introduction}

A lot of study has been devoted to the understanding of the glass
transition due to its theoretical, fundamental and technological importance
\cite{angell}. These meta-stable states can be found in both simple
and colloidal fluids\cite{debenedetti}. Colloidal systems offer,
however, the unique opportunity to tailor the interaction potential
between the particles, and the time scale, being much longer, allows
an easier study of such systems, from the experimental stand point
\cite{mallamace,chen1,chen2,grandjean,pontoni,pham,pham2}.

The foremost theoretical formalism to account for much of the experimental
work done so far is the Mode-Coupling (MC) approach \cite{goetze1}. One
of the most impressive predictions of this theory was the glass-liquid-glass
re-entrant transition in hard-spheres colloidal fluids \cite{bergenholtz,fabian}.
The theory nonetheless yields conflicting answers when a bidisperse
system \cite{pham,pham2} is treated as it is, i.e. a true bidisperse
system, and as compared with an effective one-component fluid under
the MC approach \cite{zaccarelli}. Recently, however, a theory that
seems to be better suited to the study of these systems, mainly in
multicomponent case, has been put forward \cite{rigo1,rigo2}.
We henceforth refer to it as the self-consistent generalized langevin
equation (SCGLE) theory.

On the hand of simulations experiments there is several attempts of
investigate the static and dynamic properties of hard sphere mixtures.
Particullarly have made studies of the dynamics of polidisperse systems,
by however, none has achieved the kinetic phase diagrams.

The importance of the study of colloidal mixtures is crutial to the understanding
of the dynamic properties of polidisperse systems, because there is not
monocomponent ideal system in the nature. Especialy, for simplicity, the case of
two and three species is fundamental to understand the more basic properties
of such systems. Specifically the clasification of the different types of phases
that may occur is fundamentally important. That classification can be drawn
in what is known as kinetic phase diagram.
Particularly the topology of the diagram such as we will see, say a lot about the different
states of the system.
 
In this work, we use SCGLE theory to analyze the likelihood of getting a second
"re-entrant" transition when a smaller particle species is introduced 
into a bidisperse system. Thus the system that
we are envisioning is a three component mixture of hard spheres
with size asymmetry, namely, hard spheres of relative sizes $1$, 
$3$ and $9$, which we will refer to as $1:3:9$.
And with help of the localization length we can catalog the differents
types of phases of the system.

\section{SCGLE theory for dynamic arrest}
The SCGLE theory of dynamic arrest was introduced in Refs. \cite{rmf, todos1, todos2, attractive1} and extended
to colloidal mixtures in Ref. \cite{rigo1}. 
The relevant dynamic information of an equilibrium 
$\nu$-component colloidal suspension is contained in the $\nu \times \nu$ matrix $F(k,t)$ 
whose elements are the {\it partial intermediate scattering functions} 
$F_{\alpha \beta}(k,t)\equiv \left\langle n_{\alpha}({\bf k},t)n_{\beta}(-{\bf k}^{\prime},0)\right\rangle$, where $n_{\alpha}({\bf k},t)\equiv \sum_{i=1}^{N_{\alpha}}
\exp[i{\bf k}\cdot {\bf r}_i(t)]/\sqrt{N_\alpha}$, with ${\bf r}_i(t)$ being the position of particle $i$ of
species $\alpha$ at time $t$. The initial value $F_{\alpha \beta}(k,0)$ is the partial static structure factor
$S_{\alpha \beta}(k)$ \cite{HANSEN, NAGELE}. The SCGLE theory is summarized by a self-consistent system of
equations \cite{rigo1,marco2} for the $\nu \times \nu$ matrices $F(k,t)$ and $F^{(s)}(k,t)$. The latter being defined
as $F_{\alpha\beta}^{(s)}(k,t)\equiv \delta_{\alpha \beta}\left\langle \exp {[i{\bf k}\cdot \Delta{\bf
R^{(\alpha )}}(t)]} \right\rangle $, where $\Delta{\bf R}^{(\alpha )}(t)$ is the displacement of any of the
$N_{\alpha }$ particles of species ${\alpha}$ over a time $t$, and $\delta_{\alpha \beta}$ is Kronecker's delta
function. As illustrated in Ref. \cite{rigo1}, the solution of the SCGLE theory above provides the time and
wave-vector dependence of the main dynamic properties. It also provides equations for their long-time asymptotic
values, referred to as non-ergodicity parameters, which play the role of order parameters for the
ergodic--non-ergodic transitions. The most fundamental of these results \cite{rigo1} is the following equation
for the asymptotic mean squared displacement $\gamma_\alpha \equiv \lim_{t \to \infty} <(\Delta
\textbf{R}^{(\alpha)})^2>$,

\begin{equation}
\begin{split}
\frac{1}{\gamma_{\alpha}}=\frac{1}{3(2\pi)^3}\int d^3k k^2 &
\left\lbrace\lambda [ \lambda  +k^2\gamma ]
^{-1}\right\rbrace_{\alpha\alpha} \\
& \times \left\lbrace c \sqrt{n} S \lambda [ S \lambda +k^2\gamma ] ^{-1} \sqrt{n} h\right\rbrace
_{\alpha\alpha},
\end{split}
\label{1}
\end{equation}
where $S$ is the matrix of partial static structure factors, $h$ and
$c$ are the Ornstein-Zernike matrices of total and direct
correlation functions, respectively, related to $S$ by $S =
I+\sqrt{n}h\sqrt{n} = [I-\sqrt{n}c\sqrt{n}]^{-1}$, with  the matrix
$\sqrt{n}$ defined as $[\sqrt{n}]_{\alpha\beta} \equiv
\delta_{\alpha\beta}\sqrt{n_\alpha}$, and  $\lambda(k)$ is a
diagonal matrix given by $\lambda_{\alpha\beta} (k) = \delta_{\alpha
\beta} [1+(k/k^{(\alpha)}_c)^2]^{-1}$, where $k^{(\alpha)}_c$ is the
location of the first minimum following the main peak of $S_{\alpha
\alpha}(k)$( a more practical rule is to use simply $k^{(\alpha)}_c=1.545 (2\pi/{\sigma_{\alpha}})$,
 with $\sigma_{\alpha}$ the diameter of species $\alpha$).
The solution of Eq. \ref{1} give us the criterion to determine
when a species is arrested or not, namely, if $\gamma_{\alpha} \to \infty$ 
the species $\alpha$ is in a fluid state and if $\gamma_{\alpha}$ is finite then
it is in arrested state; particularly $\sqrt{\gamma_{\alpha}}$ is its localization length.

\section{The system description}
Let a three component hard-sphere-like colloidal system be composed of $N$ particles
with three different species (A,B,C) of which $N_A$ are of diameter $\sigma_A$, 
$N_B$ of $\sigma_B$ and $N_C$ of $\sigma_C$ and $N=N_A+N_B+N_C$. For simplicity,
we will consider the case where $\sigma_A=9$, $\sigma_B=3$ and $\sigma_C=1$.
The control parameters will be the volume fractions $\phi_A$,$\phi_B$
and $\phi_C$ of the species A, B and C respectively; where $\phi_{\alpha}=\pi n_{\alpha}\sigma_{\alpha}^3 /6 $,
with $n_{\alpha}=N_{\alpha}/V$ and $V$ is the volume of the system.
The size asymmetry parameter $\delta_{(ij)}=\sigma_i/\sigma_j$ can be useful.
We have numerically resolved the Eq. (\ref{1}) for the above system and calculated
its full dynamic arrest phase diagram. In calculation of the matrix
of partial static structure factors needed to resolve the Eq. (\ref{1})
we have used the analytical Baxter \cite{baxter} solution of Ornstein-Zernike equations
with Percus-Yevick approximation for a multicomponent hard sphere system.

\section{Results}
To describe the dynamic arrest transitions of a ternary hard sphere mixture we need a three dimensional
cartesian space, is to say, the space $(\phi_A,\phi_B,\phi_C)$. The results are as follows:

\subsection{The $(\phi_A,\phi_B)$ plane}

\begin{figure}[h]
\begin{center}
\includegraphics[scale=0.3]{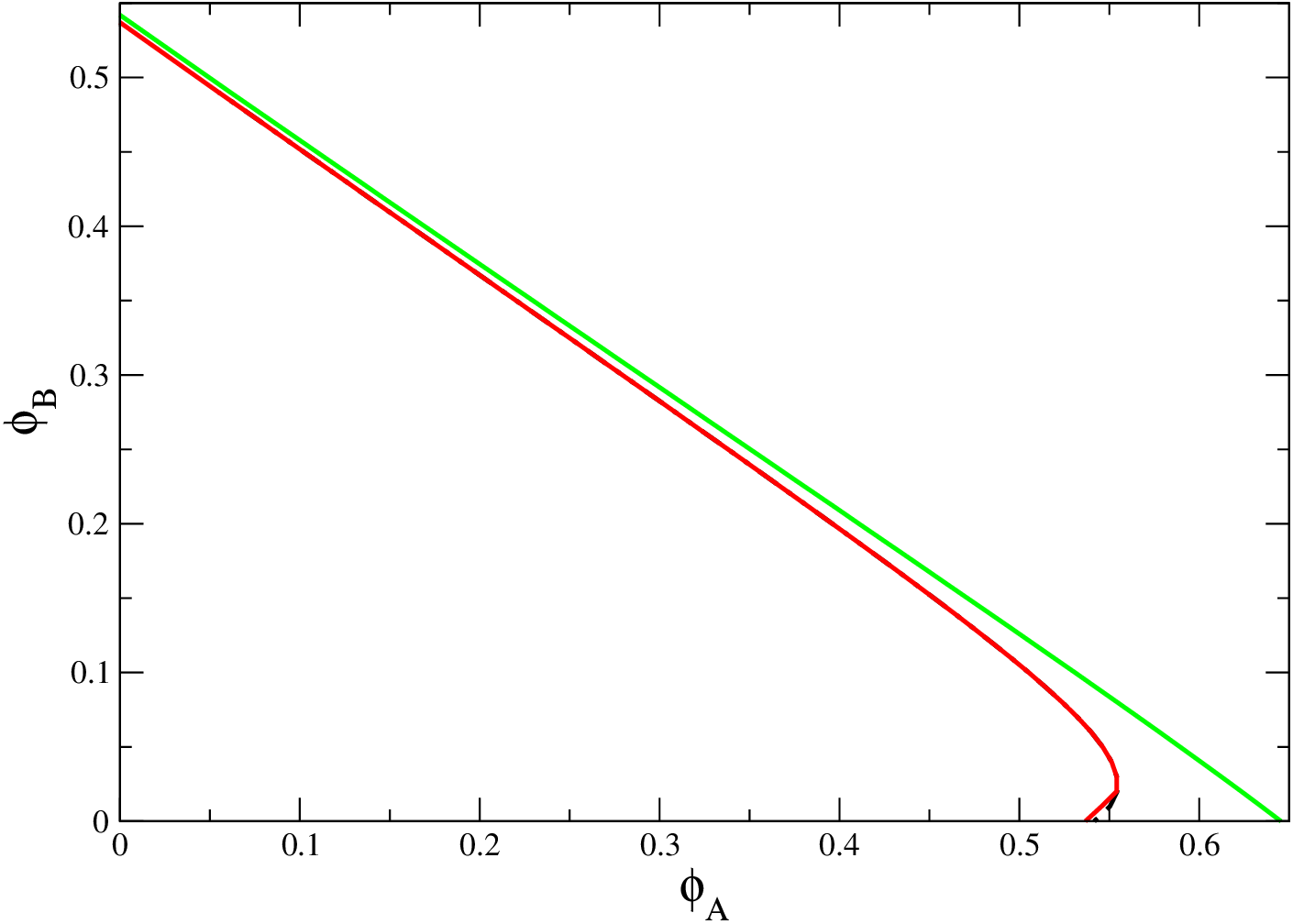}
\caption{Dynamic arrest phase diagram of a ternary hard sphere mixture
 with $\phi_C=0$. The red line represents the transition of species A from fluid state to an arrested state, 
the dotted line indicates the virtual localization of species B, and the green dashed line
corresponds to the localization transition of particles C}\label{fig1}
\end{center}
\end{figure}

We start by showing the dynamic arrest phase diagram (Fig. \ref{fig1}) for the case in which the
species C (smaller) is found at infinite dilution.
The red curve represents the line of arrest for
the large particles; below and the left of it the large particles
are found in liquid state, while above of it lies the region of arrested states. 
The interception of this line with
the $\phi_A$ axis represents the arrest of the large particles when there
are no other particles (monocomponent glass transition point), 
which happens at the volume fraction of $\phi_{A}=0.537$.
As we increase the volume fraction of species B, the lines
move to the right due to the effect of depletion forces as discussed
elsewhere \cite{asakura, rigo2}. The red line moves to approximately $\phi_A=0.554$
where it joints the dashed black line (line of arrest of the middle species, see zoom Fig. \ref{fig2}),
and after that both lines move together to the left, making a curve of simultaneous arrest of species A and B.
Between the red curve and the black curve we find
mixed states where particles A remain arrested but particles B
are moving in an ergodic state.
The nature of this hybrid states can be undersood because the species A reached the ideal glass transition, 
and the presence of species B and C (which continues to spread) only shifts the transition point.
As we increase the volume fraction of species A, the accessible volume to the species 
B is getting smaller until eventually it is located. 
We also show a green line, representing the line of arrest for the
small particles, which are found at infinite dilution through out
this graph, i.e., there remains just a trace of them. Thus above the green
curve and to its right all particles are found in the glass, non-ergodic
state, while between lines we have corresponding mixed states.
The crossings of these lines with the $\phi_A$ axis represent the sequential
arrest of the A, B and C species found at the following
volume fractions of the third species: $\phi_A=0.537$, $0.542$  and $0.645$.
If we disregard the green line (the line of arrest for species C),
what we get is the same kind of diagram reported
elsewhere for binary hard-sphere mixture
\cite{rigo2}. In contrast with the two component system, we now have
the possibility of having two kinds of arrested species, A and B,
and the C species moving in the porous medium formed by the matrix of obstacles created by
the arrest of species A and B. This is the region of states between the green and red curves.

\subsection{First re-entrance}

\begin{figure}[h]
\begin{center}
\includegraphics[scale=0.25]{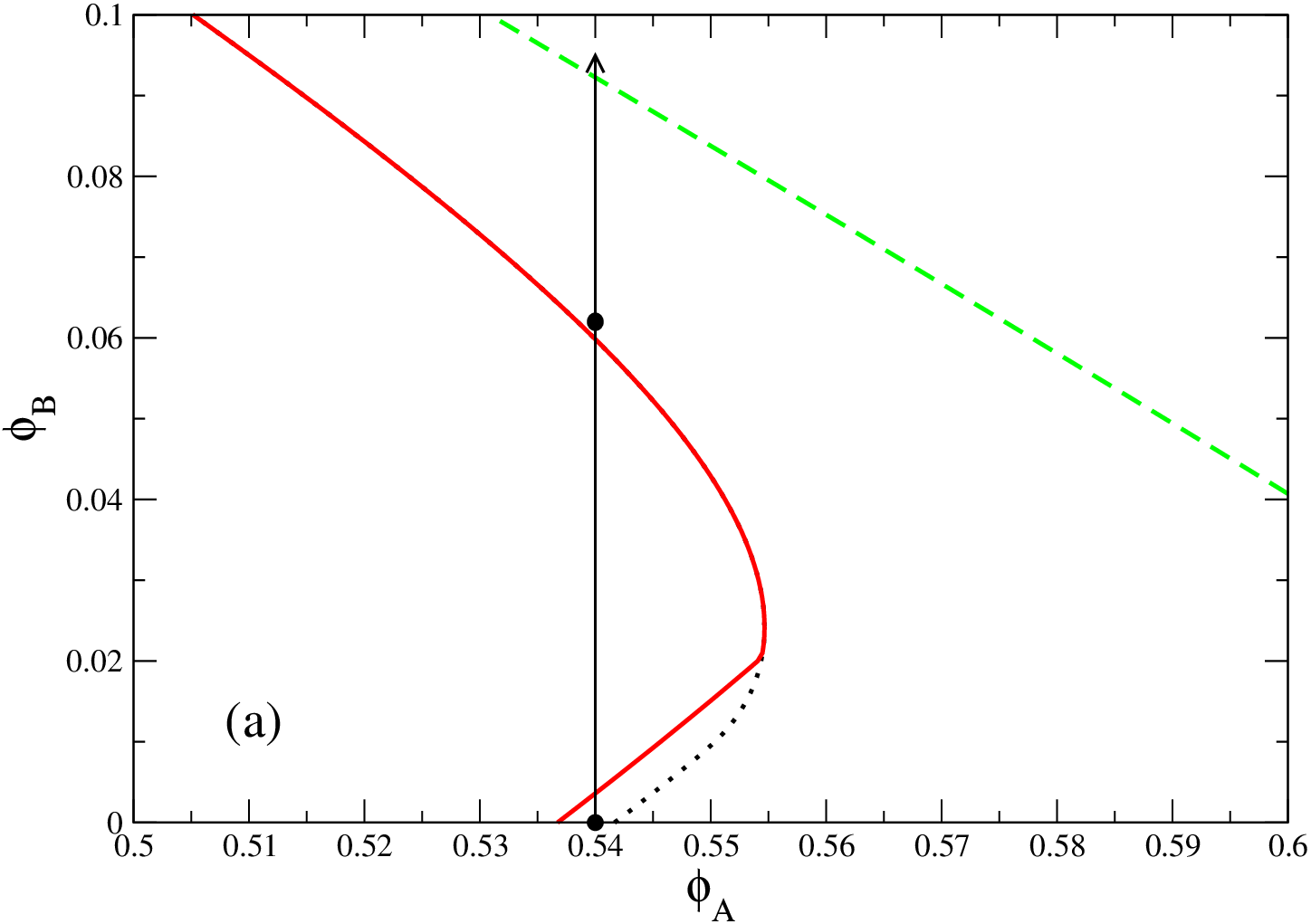}
\includegraphics[scale=0.25]{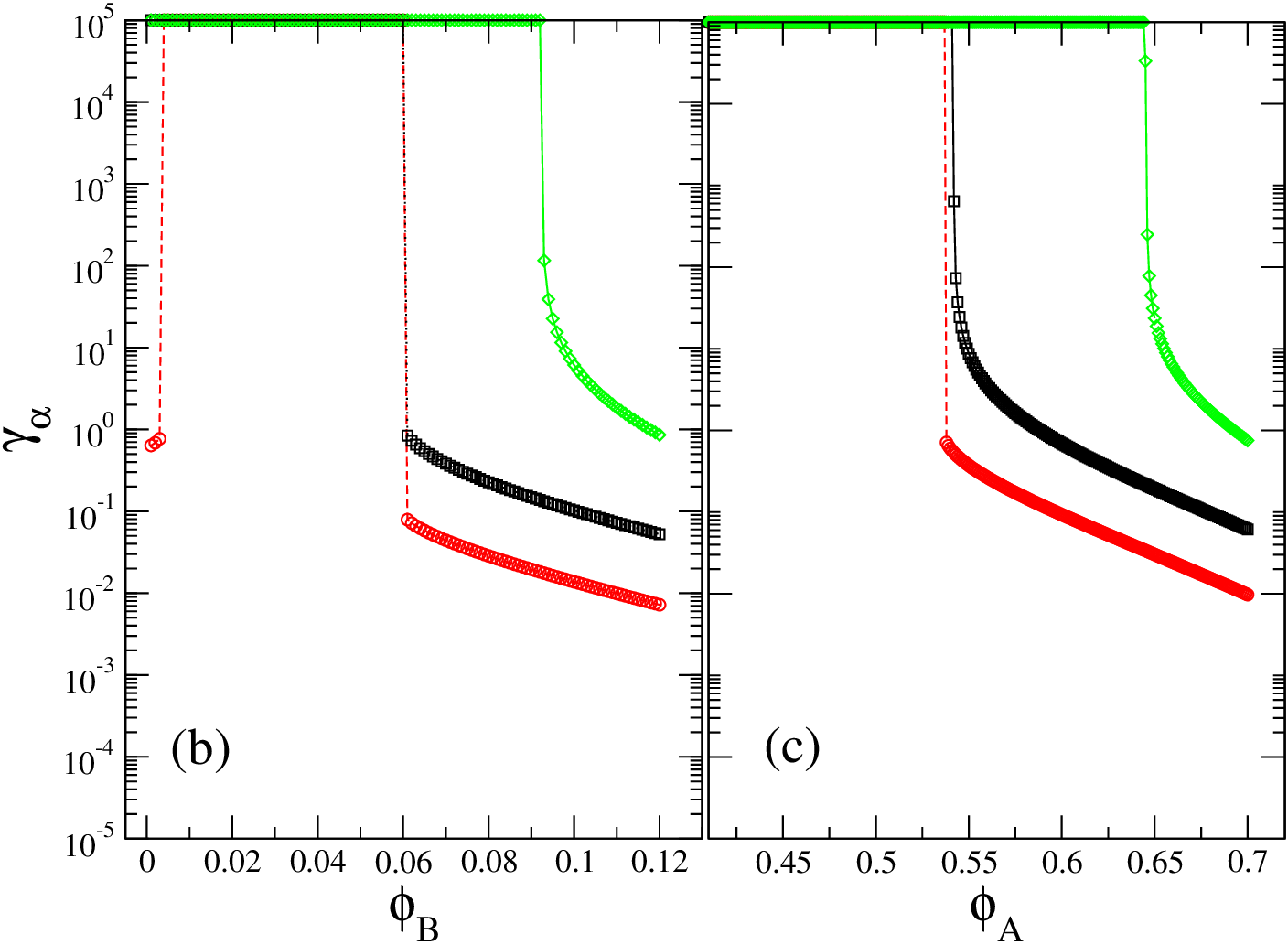}
\caption{(a) This is a close up of Fig. \ref{fig1}. 
The lower point represent a mixed state where
paricles A are arrested while the B and C are moving.
The upper point is an state where the particles A and B are arrested
while the C are diffusing. (b) Shows the square of localization length
of specie A (red), specie B (black) and specie C (green); calculated along
the arrow indicated in Fig2a. (c) Square of localization length but now along
the axis $\phi_A$, i.e., with only a trace of species B and C. In all cases
the units of localization length of specie $\alpha$ is its own diameter $\sigma_{\alpha}$.
And the top numeric label ($10^5$) of vertical axis in (b) and (c) is a representation of infinity,
i.e., the localization length is infinite (particles in a fluid state.) 
}\label{fig2}
\end{center}
\end{figure}

An interesting phenomenon already discussed in the case of the bidisperse
system is the re-entrance in the kinetic phase diagram, whereby we mean the following: consider
a two-component system whose state is represented by a point in the
$\phi_A$ axis at volume fraction of, say, $\phi_A=0.54$; at this volume fraction
the species A is already arrested (lower dot in Fig. \ref{fig2} ).
If we now increase the volume fraction of the species B, we move upwards
in this phase diagram. By doing this, we enter to the
ergodic zone, specifically, the B particles are found in the liquid state.

This effect is due to the depletion forces caused by the B particles, 
whereby the A particles feel an attraction of the
Asakura-Oosawa \cite{asakura} type and get closer together, which makes room for
the particles to move; in other words, since the cages
open up, the system melts. Not only is there more room for the large
particles but for all particles as well, as can be inferred by the
fact that all lines start moving to the right.
If we still continue to increase the volume fraction of 
particles B, we arrive at a total volume fraction in which both species
get arrested, this is represented by the upper dot in Fig. \ref{fig2}.
From the point of intersection, both lines (black and red) merge
into one up to the left extreme where $\phi_A=0$, here the arrest
of the species B arrests the large ones, since there are no holes
through which they can go through. The particles C, however, still have
some room and at this extreme they get arrested when $\phi_B=0.542$.
As we can see from Fig. \ref{fig2}, there is a gap between the green and
red-black lines that do not intersect even when the volume fraction of
the species A (the larger particles) vanishes. And what happens
is this: when we have no A and C species, the
species B gets arrested at the volume fraction of $\phi_B=0.537$, and the trace
of species A is arrested by the particles B. The species C gets arrested
at the expected value of $\phi_C=0.542$, since at this end on the phase
diagram, where $\phi_C=0$, the system is equivalent to a bidisperse
system with a size asymmetry of $\delta_{(BC)}=\sigma_B/\sigma_C=0.333$; 
then we have to recover
the same values of arrest found in the $\phi_A$ axis by the black(arrest of particles B) and
red lines(arrest of particles A), since $\delta_{(AB)}=\delta_{(BC)}=0.333$.

On the other hand, at small volume fractions of the species B, the glass is due
to the cage effect and we call it {\it repulsive glass} \cite{pham,rigo2}. However
at higher volume fractions of it, the glass formed
by the large particles is of a different type. This part of the diagram
is called {\it attractive glass} \cite{pham,rigo2}, since what we see, disregarding the presence
of the particles B, that there is an effective attraction among the particles of type A,
and the localization length of particles B becomes smaller
than particles B on the first part of the curve by two orders of magnitude.
If we continue along this line to smaller
volumes fractions of the particles A until only a trace of them is found
we have the "{\it chancaquilla effect}" \cite{rigo1}, where the particles A are
arrested because the particles B get arrested. For intermediate
volume fractions of the particles A, they get arrested because
of the combined effect of the volume fractions; that is, due to the total volume
fraction (at this size asymmetry).

\begin{figure}[h]
\begin{center}
\includegraphics[scale=0.3]{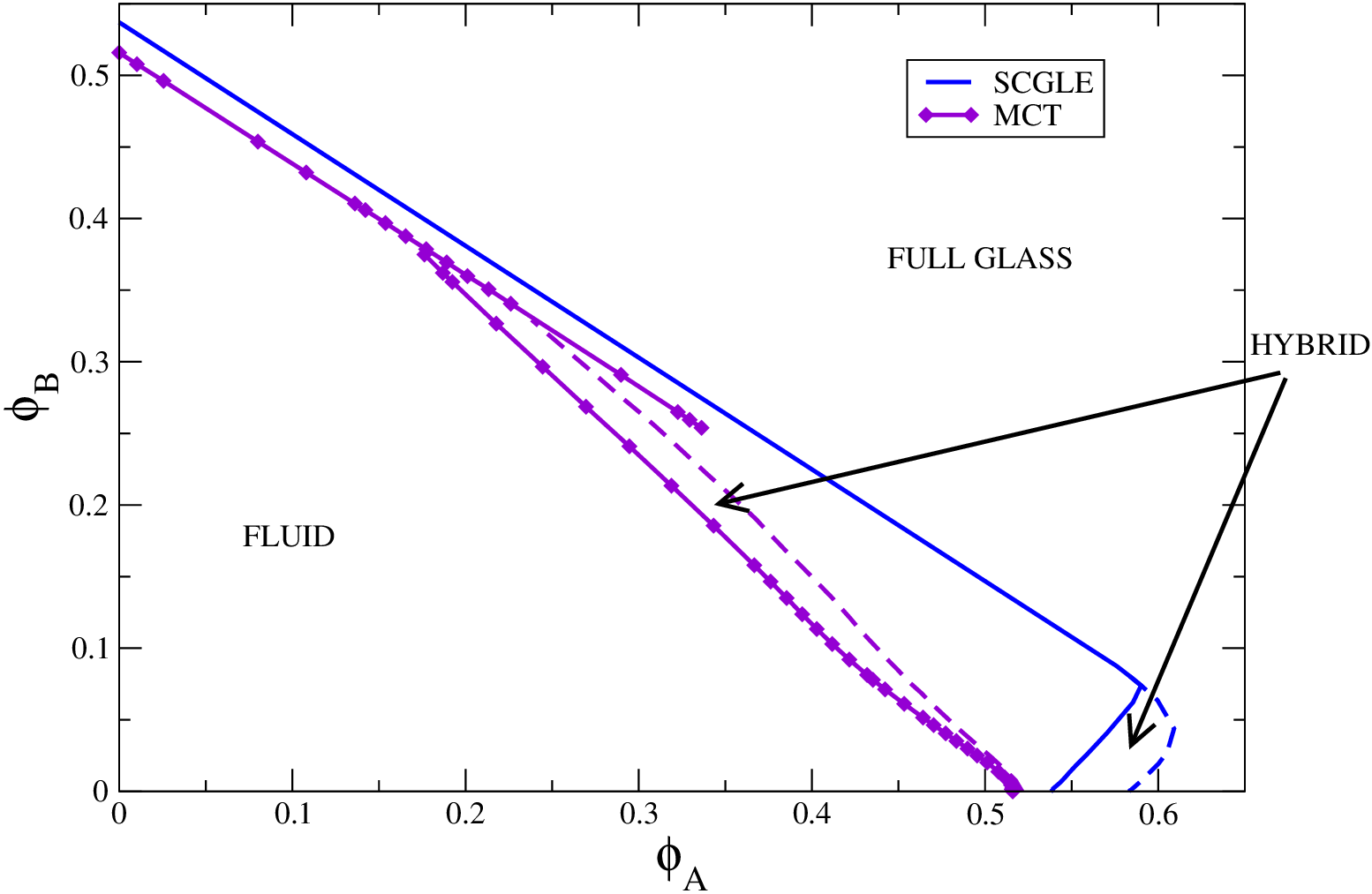}
\caption{Cuantitave comparison between the dynamic arrest diagrams
of MCT theory (violet symbol-line) and SCGLE theory (blue line). 
In both diagrams are distinguished three region: fluid, full glass
and partial (hybrid) glass states .
Note that MCT theory do not show reentrant effect.}\label{mct}
\end{center}
\end{figure}

\subsection{Second re-entrance}

\begin{figure}[h]
\begin{center}
\includegraphics[scale=0.3]{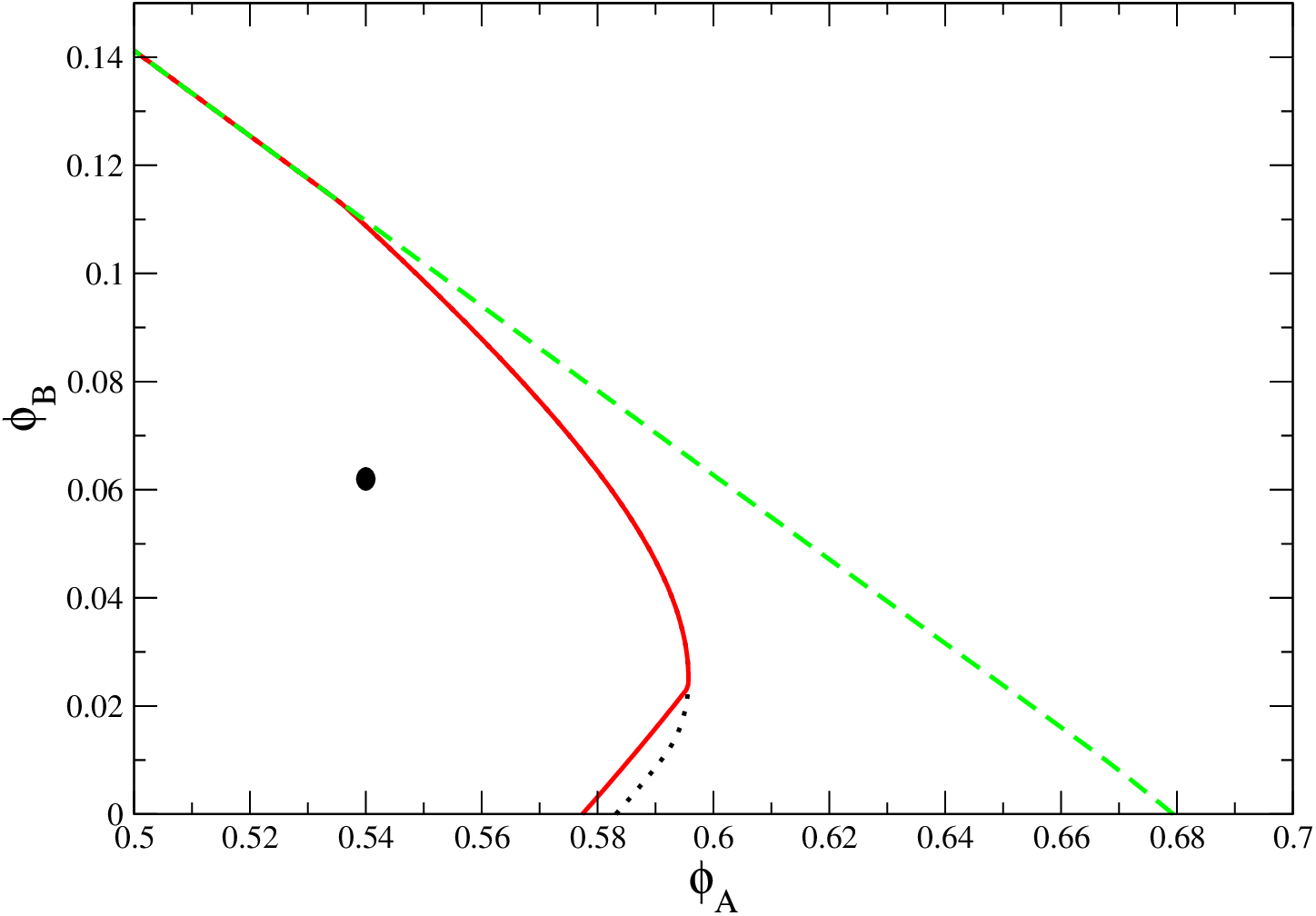}
\caption{Here we show the $\phi_C=0.05$ plane. Note that the arrest lines moved
to the right, and this time all lines merge at some point. The dot
shown is inside the ergodic region and correspond to the dot
in Fig. \ref{fig2}.}\label{fig3}
\end{center}
\end{figure}

\begin{figure}[h]
\begin{center}
\includegraphics[scale=0.3]{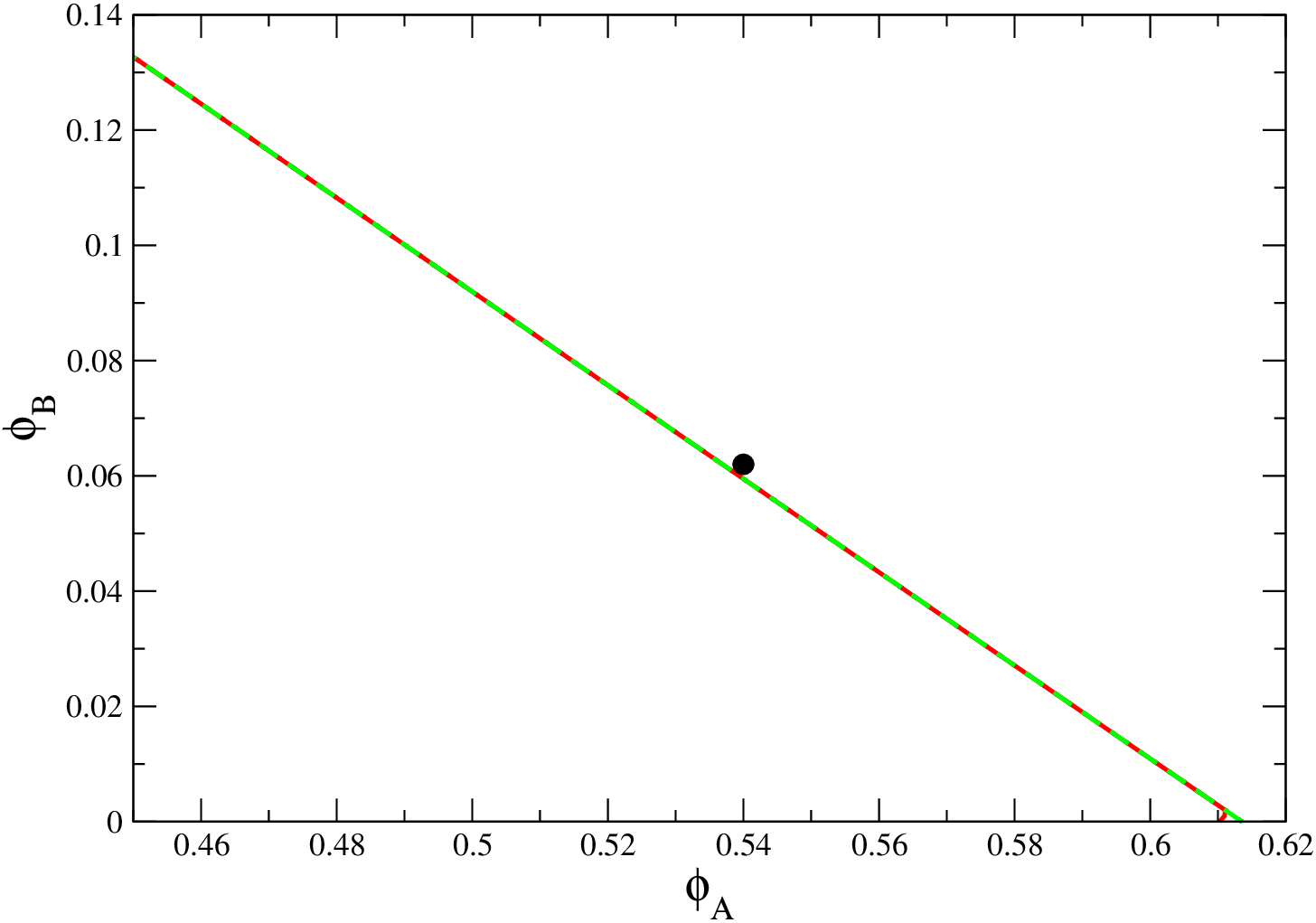}
\caption{This is the $\phi_C=0.1$ plane. Note that all arrest lines have merged
into one. The black dot finds itself in the non-ergodic region again.}\label{fig4}
\end{center}
\end{figure}

In the discussion of previous section, the species C has played no active role
since it was found at infinite dilution. Our system was essentially a
two component. However, we have three species, so the phase 
diagram is actually tridimensional. Up to now, we have shown only the $(\phi_A,\phi_B)$ axis slice
of the three dimensional phase space when $\phi_C=0$.
We will now increase the volume fraction of species C, i.e., we
will move along the $\phi_C$ axis and show different slices at different
values of the volume fraction of the particles C. 
The next slice that we show corresponds to $\phi_C=0.05$ in Fig. \ref{fig3}.
Note that the lines move to the right, once more as a consequence of depletion
forces. This causes a second re-entrance, but this time due to the addition
of particles C. The black dot has the same coordinates as that
in Fig. \ref{fig2} and, as we can see, this point finds itself in the
ergodic region. Note also that this time the arrest lines of particle A
and B merge at around $\phi_3=0.595$.
And if we continue increasing the volume fraction of species C, we
do indeed get arrested again, Fig. \ref{fig4}.  This time all particles are
arrested, as we can see. Here the three lines have merged into
one, and the gap has vanished. The black dot is located once again in the non-regodic region.

\subsection{The $(\phi_A,\phi_C)$ plane}

\begin{figure}[h]
\begin{center}
\includegraphics[scale=0.25]{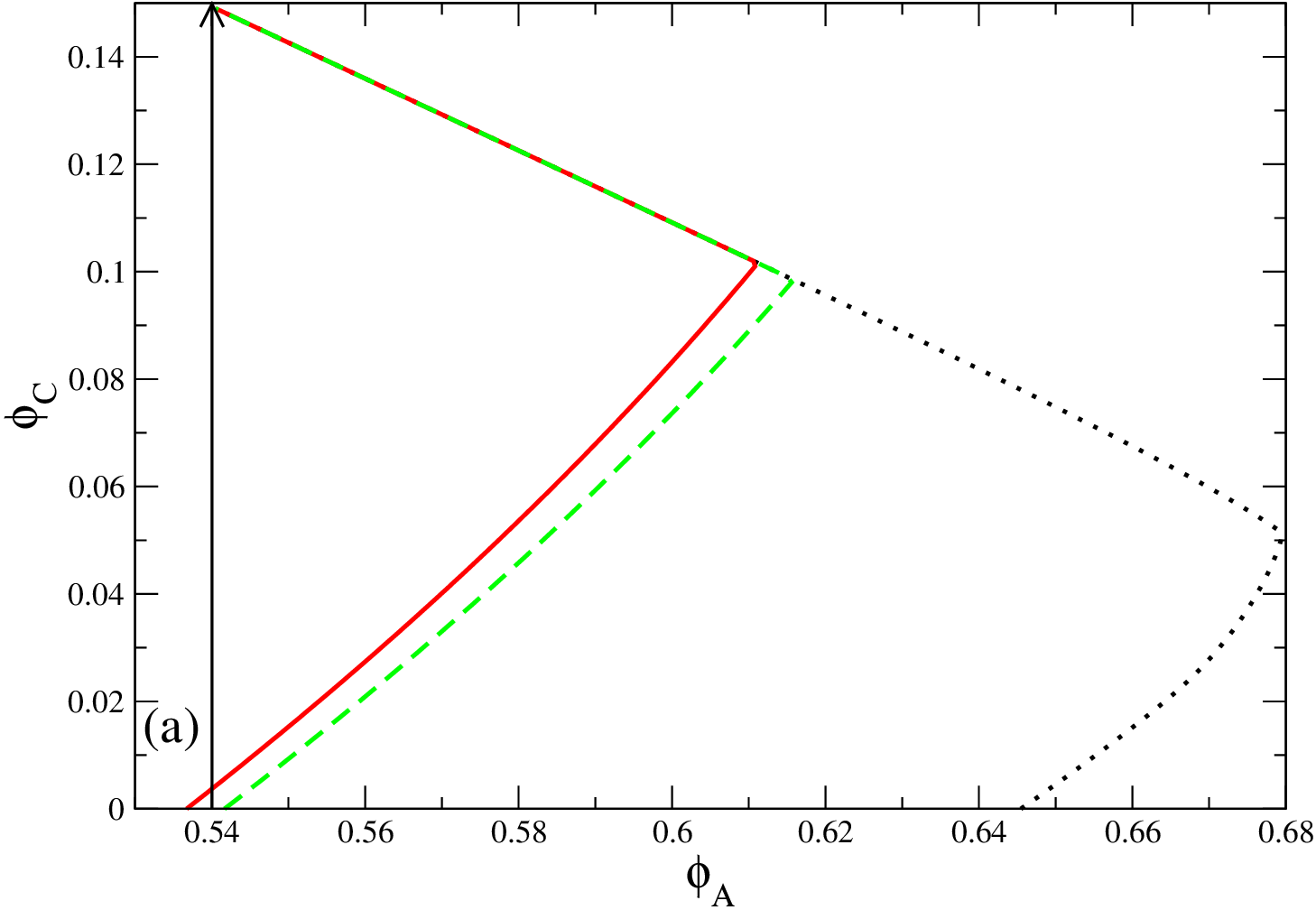}
\includegraphics[scale=0.25]{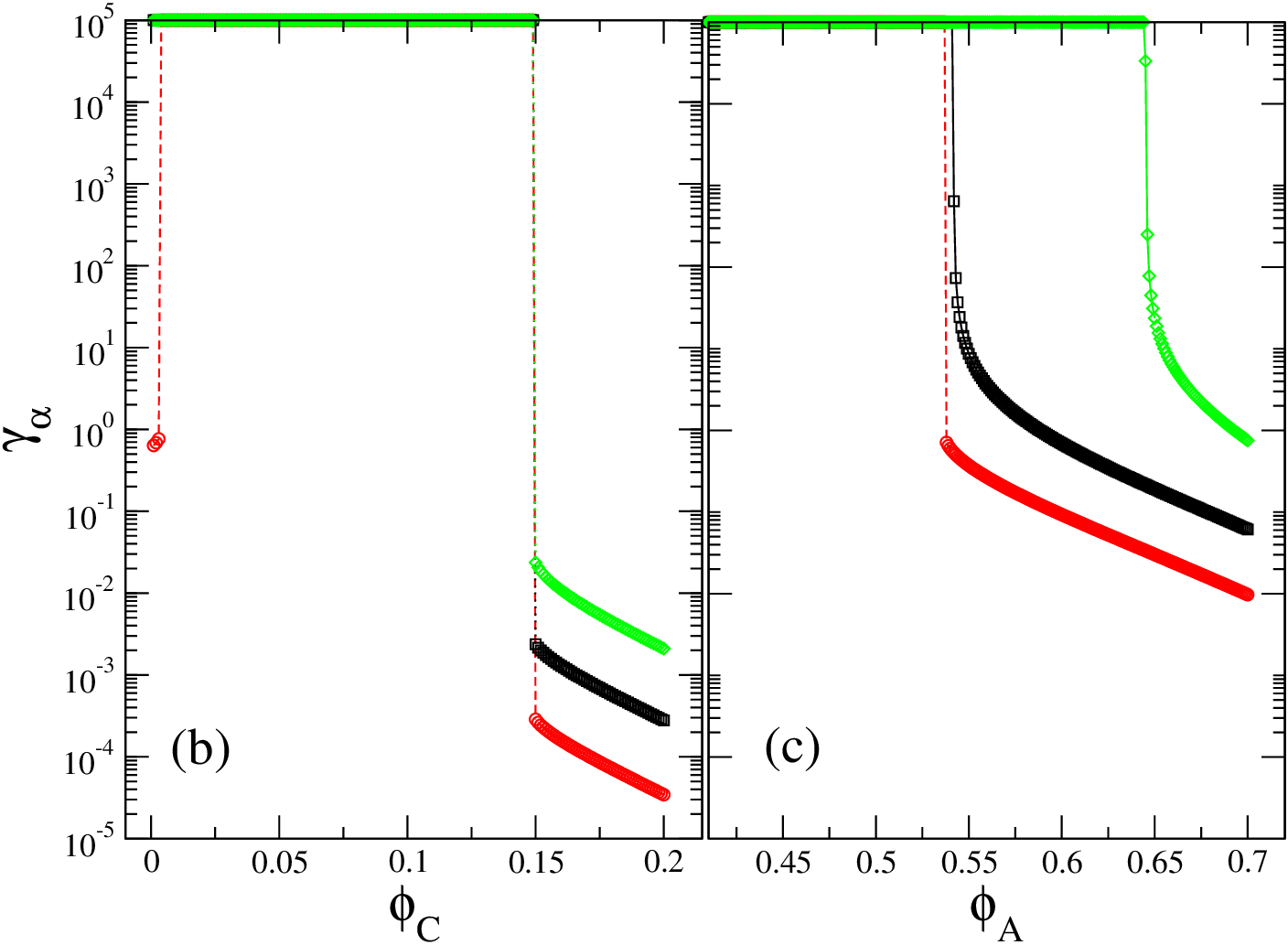}
\caption{(a) Plane $\phi_B = 0.0$, it shows the arrest lines of species A (red line),
B (dashed green line) and C (black dotted line). (b) and (c) figures show the square localization
lengths of particles of type A (red), B (black) and C (green) along the arrow drawn in figure (a) and along the
$\phi_A$ axis, respectively. The units are the same of Fig 2.}\label{fig5}
\end{center}
\end{figure}

Now we will take a look at the $(\phi_A,\phi_C)$ plane, where the
particles B are found at infinite dilution (see Fig. \ref{fig5}).
If we disregard the green dashed curve, which corresponds to the line of
arrest of the particles B, the remaining curves correspond to a binary system with an asymmetry of
$\delta_{CA}=\sigma_C/\sigma_A=0.111$.
This situation is quite analogous to the $(\phi_A,\phi_B)$ plane of Figure \ref{fig2}, provided
the same green dashed line is once more disregarded. Only the effect of the asymmetry is
more important. Also, in a similar manner as we had before, there
are two kinds of glasses made by the large particles. In the lower
part of the red curve, the large particles form a repulsive glass
whereas in the upper part they form an attractive glass. This time,
however, the localization length of the large particles drops by five
orders of magnitude.
Another line of arrest, but not less important, corresponds to the
localization of species B (dashed green line). 
It is worthwhile noting that although there is only one trace of them, 
their phase diagram is governed or strongly correlated with the species A 
(continued red line).

\section{Conclusions}
In summary we have investigated the dynamic arrest phase diagrams
of a ternary hard sphere mixture with relative sizes 1:3:9, and found that
there is a glass-fluid-glass-fluid-glass transition of the bigger species.
The first glass-fluid-glass is due the depletion forces caused by
the presence of the middle particles and the second re-entrance 
as a result of adding an even smaller species. The full kinetic phase diagrams
for the cases $\phi_C=0.0$, $\phi_C=0.05$, $\phi_C=0.1$ and $\phi_B=0$ were
shown.
We also shown the differences between MCT and SCGLE theories in the kinetic
phase diagrams of a binary mixture of hard spheres with size asymmetry of 0.5.
These differences are highly relevant for example when it is required to describe 
the properties of a polymer colloid mixture.
And we saw clearly that the MCT theory is not consistent with such experimental results,
for instance the best and most consistent description the SCGLE theory.
In light of these results we conclude that the SCGLE theory is an excellent
tool when we will try to describe systems with a high degree of polydispersity.


ACKNOWLEDGMENTS: This work was supported by PROMEP under grant
PROMEP-103.5-11-4911.

\end{document}